\documentclass[twocolumn]{aastex62}

\usepackage[shortcuts]{extdash}

\def\apjl{ApJL}

\def\aap{A\&A}

\def\apjs{ApJS}

\def\swift{\textit{Swift}}
\def\chandra{\textit{Chandra}}
\def\xmm{\textit{XMM-Newton}}
\def\nustar{\textit{NuSTAR}}

\newcommand{\fluxcgs}{ergs~s$^{-1}$~cm$^{-2}$}
\newcommand{\lumcgs}{ergs~s$^{-1}$}
\newcommand{\msol}{$M_{\odot}$}

\received{}
\revised{}
\accepted{}
\submitjournal{ApJ}

\shorttitle{Variability of NGC 925 ULX-3}
\shortauthors{Earnshaw et al.}

\begin{document}

\title{The variability behavior of NGC 925 ULX-3}

\correspondingauthor{Hannah Earnshaw}
\email{hpearn@caltech.edu}

\author[0000-0001-5857-5622]{Hannah P. Earnshaw}
\affil{Cahill Center for Astronomy and Astrophysics, California Institute of Technology, Pasadena, CA 91125, USA}

\author[0000-0002-8147-2602]{Murray Brightman}
\affil{Cahill Center for Astronomy and Astrophysics, California Institute of Technology, Pasadena, CA 91125, USA}

\author{Fiona A. Harrison}
\affil{Cahill Center for Astronomy and Astrophysics, California Institute of Technology, Pasadena, CA 91125, USA}

\author[0000-0002-1082-7496]{Marianne Heida}
\affil{European Southern Observatory, Karl-Schwarzschild-Strasse 2, 85748, Garching bei M\"{u}nchen, Germany}

\author[0000-0002-3850-6651]{Amruta Jaodand}
\affil{Cahill Center for Astronomy and Astrophysics, California Institute of Technology, Pasadena, CA 91125, USA}

\author[0000-0002-8183-2970]{Matthew J. Middleton}
\affil{Department of Physics \& Astronomy, University of Southampton, Southampton, SO17 1BJ, UK}

\author[0000-0001-8252-6337]{Timothy P. Roberts}
\affil{Centre for Extragalactic Astronomy \& Department of Physics, Durham University, South Road, Durham DH1 3LE, UK}

\author[0000-0001-5819-3552]{Dominic J. Walton}
\affil{Centre for Astrophysics Research, University of Hertfordshire, College Lane, Hatfield AL10 9AB, UK}
\affil{Institute of Astronomy, University of Cambridge, Madingley Road, Cambridge CB3 0HA, UK}

\begin{abstract}

We report the results of a 2019--2021 monitoring campaign with \swift\ and associated target-of-opportunity observations with \xmm\ and \nustar, examining the spectral and timing behavior of the highly variable ultraluminous X-ray source (ULX) NGC~925~ULX-3. We find that the source exhibits a 127--128 day periodicity, with fluxes typically ranging from $1\times10^{-13}$ to $8\times10^{-13}$\,\fluxcgs. We do not find strong evidence for a change in period over the time that NGC~925~ULX-3 has been observed, although the source may have been in a much lower flux state when first observed with \chandra\ in 2005. We do not detect pulsations, and we place an upper limit on the pulsed fraction of $\sim$40\% in the \xmm\ band, consistent with some previous pulsation detections at low energies in other ULXs. The source exhibits a typical ULX spectrum that turns over in the \nustar\ band and can be fitted using two thermal components. These components have a high temperature ratio that may indicate the lack of extreme inner disk truncation by a magnetar-level magnetic field. We examine the implications for a number of different models for superorbital periods in ULXs, finding that a neutron star with a magnetic field of $\sim$10$^{12}$\,G may be plausible for this source. The future detection of pulsations from this source would allow for the further testing and constraining of such models. \\

\end{abstract}

\section{Introduction} \label{sec:intro}

Ultraluminous X-ray sources (ULXs) are non-nuclear X-ray sources with luminosities above $\sim$10$^{39}$\,erg\,s$^{-1}$, widely agreed at this time to be a population dominated by stellar-mass compact objects accreting at super-Eddington rates (for recent reviews, see \citealt{kaaret17,fabrika21}). Many ULXs are persistently bright but a subset show extreme variability, sometimes of over an order of magnitude in flux, in which they may even enter and leave the super-Eddington luminosity regime. Some of these sources appear to be one-off excursions into the ULX regime, such as particularly bright instances of classical outbursts (e.g. \citealt{middleton13}), or other hard and bright transient ULXs with less certain interpretations (e.g. \citealt{earnshaw19,earley21}). However, many spend significant amounts of their duty cycle in the ULX luminosity regime, or exhibit high-amplitude variability while remaining at ULX luminosities throughout. 

One proposed mechanism for such extreme variability in ULXs is the propeller effect (e.g. \citealt{tsygankov16} for M82 X-2, but see below), in which accretion is halted when the magnetospheric radius of the accreting neutron star exceeds the corotation radius of the accretion disk \citep{illarionov75,stella86}. This would cause the flux to drop dramatically but not necessarily regularly. 

Another feature of high-amplitude variability found in some ULXs is (quasi-)periodic variability with timescales of tens to hundreds of days (e.g. \citealt{strohmayer09,walton16,vasilopoulos20}). In some cases, this is distinct from a known orbital period and attributed to superorbital periodicity. Detection of this kind of long-term periodicity requires monitoring over long timescales, and sometimes can be found to be the cause of variability that at first looked like the propeller effect (e.g. \citealt{brightman19} on M82 X-2). Additionally, some sources demonstrate both long-term periodicity and dramatic flux drop-outs, suggesting that both processes can potentially contribute to the long-term variability of a single source (e.g. \citealt{walton15a,walton16,israel17a,fuerst17}). In cases where the variation is confirmed to be superperiodic, the cause of the variability is not yet fully understood, but may be related to the precession within the system, and a variety of models have been proposed to explain it (e.g. \citealt{mushtukov17,middleton18,middleton19,vasilopoulos20}).

Interestingly, many known ULX pulsars discovered to date show some form of high-amplitude flux variability, and searching for such variability in archival X-ray data may be a way to identify further ULX pulsar candidates \citep{earnshaw18,song20}. However, verification of the nature of this variability requires regular monitoring over a long baseline in order to establish the magnitude of the variability and the presence of any periodicity.  

We identified NGC~925~ULX-3 as a highly variable ULX in \citet{earnshaw20}, henceforth E20. ULX\=/3 is located at 02$^h$\,27$^m$\,20$^s$.18, +33$^\circ$\,34$^\prime$\,12.84$^{\prime\prime}$ (J2000), and was serendipitously detected as a ULX in a \chandra\ observation of the spiral galaxy NGC~925 at 9.56\,Mpc. Upon searching through archival \swift\ data, we found that it had been bright on a previous occasion, but undetected or detected at significantly lower luminosities since then. Since the sparse existing temporal coverage of observations of NGC~925 made it hard to determine the nature of ULX-3's variability, we undertook a monitoring campaign using \swift, supplemented by target-of-opportunity observations with \xmm\ and \nustar\ triggered when ULX-3 was bright. 

\begin{deluxetable*}{cccc}
	\tablecaption{The observations (or ranges thereof in the case of \swift) used in this investigation. \label{tab:observations} } 
	\tablecolumns{4}
	\tablenum{1}
	\tablewidth{0pt}
	\tablehead{
		 \colhead{Observation ID} & \colhead{Observation Date} & \colhead{Instrument} & Exposure\tablenotemark{a}
	}
	\startdata
	7104 & 2005 Nov 23 & \chandra\ & 2.2 \\
	00045596001--018 & 2011 Jul 21 -- 2017 Nov 25 & \swift\ & 0.2--6.5 \\
	0784510301 & 2017 Jan 18 & \xmm\ & 37.5/37.6/27.1\tablenotemark{b} \\ 
	20356 & 2017 Dec 1 & \chandra\ & 10.0 \\
	00045596019--046 & 2019 Aug 18 -- 2020 Mar 15 & \swift\ & 0.5--3.5 \\ 
	00089002001 & 2019 Dec 13 & \swift\ & 1.9 \\
	Co-added observations\tablenotemark{c} & 2019 Nov 17 -- 2019 Dec 13 & \swift\ & 10.5 \\ 
	90501351002 & 2019 Dec 12 & \nustar\ & 53.3/52.9\tablenotemark{b} \\
	00095702001--029 & 2020 Jul 01 -- 2021 Mar 08 & \swift\ & 0.1--2.6 \\ 
	0862760201 & 2020 Aug 17 & \xmm\ & 36.8/36.9/30.5\tablenotemark{b} \\ 
	00089004001 & 2020 Aug 17 & \swift\ & 1.6 \\
    80601305002 & 2020 Aug 17 & \nustar\ & 105.8/104.9\tablenotemark{b} \\ 
    00014387001--024 & 2021 Jun 26 -- 2021 Dec 04 & \swift\ & 0.5--4.2 \\
	\enddata
	\tablenotetext{a}{The good exposure time after any filtering has been applied, in ks. Given as a range for observation ranges, and as a total for co-added observations.}
	\tablenotetext{b}{\xmm\ exposure times are given for the EPIC-MOS1/EPIC-MOS2/EPIC-pn instruments respectively, and \nustar\ times for FPMA/FPMB respectively.}
	\tablenotetext{c}{Made up of \swift\ observations 00045596035--37 co-added with 00089002001 (see Section~\ref{sec:swiftnustar}).}
\end{deluxetable*}

In this paper, we describe our observation campaign and the reduction of the X-ray data in Section~\ref{sec:data}, and the results of our analysis in Section~\ref{sec:results}, including the discovery of a 127--128-day periodicity. We note that a similar result was recently presented by \citet{salvaggio22} during the preparation of this paper. Since our analysis is slightly different and we include observations not included in \citet{salvaggio22}, we believe that this study provides a useful independent confirmation of the periodicity, as well as further spectral analysis. Finally, we present our discussion of the results in Section~\ref{sec:disc} and conclusions in Section~\ref{sec:conc}.

\section{Observations \& Data Reduction} \label{sec:data}

In order to monitor the changing brightness of NGC~925~ULX-3, we requested \swift\ Director's Discretionary Time (DDT) to observe NGC~925 on an approximately weekly basis between 2019 August 18 and 2020 March 15 (observation IDs: 00045596019--046). During that time, when the source became bright, we requested a \nustar\ DDT observation which was performed on 2019 December 12 (observation ID: 90501351002) -- since NGC~925 was not at that point visible to \xmm, there was no corresponding \xmm\ observation requested, though a \swift\ observation was performed at the same time, as is usual for \nustar\ observations (observation ID: 00089002001). We were granted further monitoring time through the \swift\ General Observer program, with observations approximately every ten days between 2020 July 1 and 2021 March 8 (observation IDs: 00095702001--029). During this monitoring period, we triggered a joint \xmm\ and \nustar\ target-of-opportunity observation during the relatively short window of \xmm\ visibility in the AO\=/19 observing cycle, when source once again reached a high-flux state. This observation (observation IDs: 0862760201 \& 80601305002) was taken on 2020 August 17, along with a \swift\ observation (observation ID: 00089004001).

Additionally, all archival \swift\ data for NGC~925 was used (observation IDs: 00045596001--018), as well as additional monitoring of NGC~925 undertaken between 2021 June 26 and 2021 December 4 (observation IDs: 00014387001--024). Results from our previous analysis of archival \xmm\ and \chandra\ data were also used, as detailed in E20. We show the observations used in Table~\ref{tab:observations}.

\subsection{Swift} \label{sec:swift}

We downloaded all \swift\ data (98 observations in total) and created clean event lists using {\tt xrtpipeline}. We extracted source and background products with the {\tt xselect} task, using a 30$^{\prime\prime}$ radius circular source region centered on the \chandra\ location of the source (E20) and a 70$^{\prime\prime}$ radius circular background region located outside the galaxy. Auxiliary response files were created using the task {\tt xrtmkarf} and the relevant redistribution matrix obtained from the CALDB. Fluxes were calculated from the background-subtracted count rate using PIMMS. As in E20, we assumed an absorbed power-law model, although this time we assumed $N_{\rm H} = 1.4\times10^{21}$\,cm$^{-2}$, based on our fits to the bright \xmm\ detection (see Section~\ref{sec:xmmnustar}), since just using the Galactic value would be an underestimate. $N_{\rm H}$ values on the order of $10^{21}$\,cm$^{-2}$ are typical for ULXs \citep{winter06}. For the power-law slope we assumed $\Gamma=1.7$ (consistent within the uncertainties for both the high-flux \chandra\ and low-flux \xmm\ observations in E20, as well as with the bright \xmm\ detection fitted by itself). Errors on the count rate (and resulting flux) were calculated in the same manner as in \citet{evans09} -- that is, for instances of $<$15 counts, the Bayesian approach of \citet{kraft91} was used to calculate the 3$\sigma$ confidence intervals. If the source was detected, the same method was used to calculate 1$\sigma$ error bars, otherwise the 3$\sigma$ upper limit was used. 

\begin{figure*}
	\begin{center}
	\includegraphics[width=18cm]{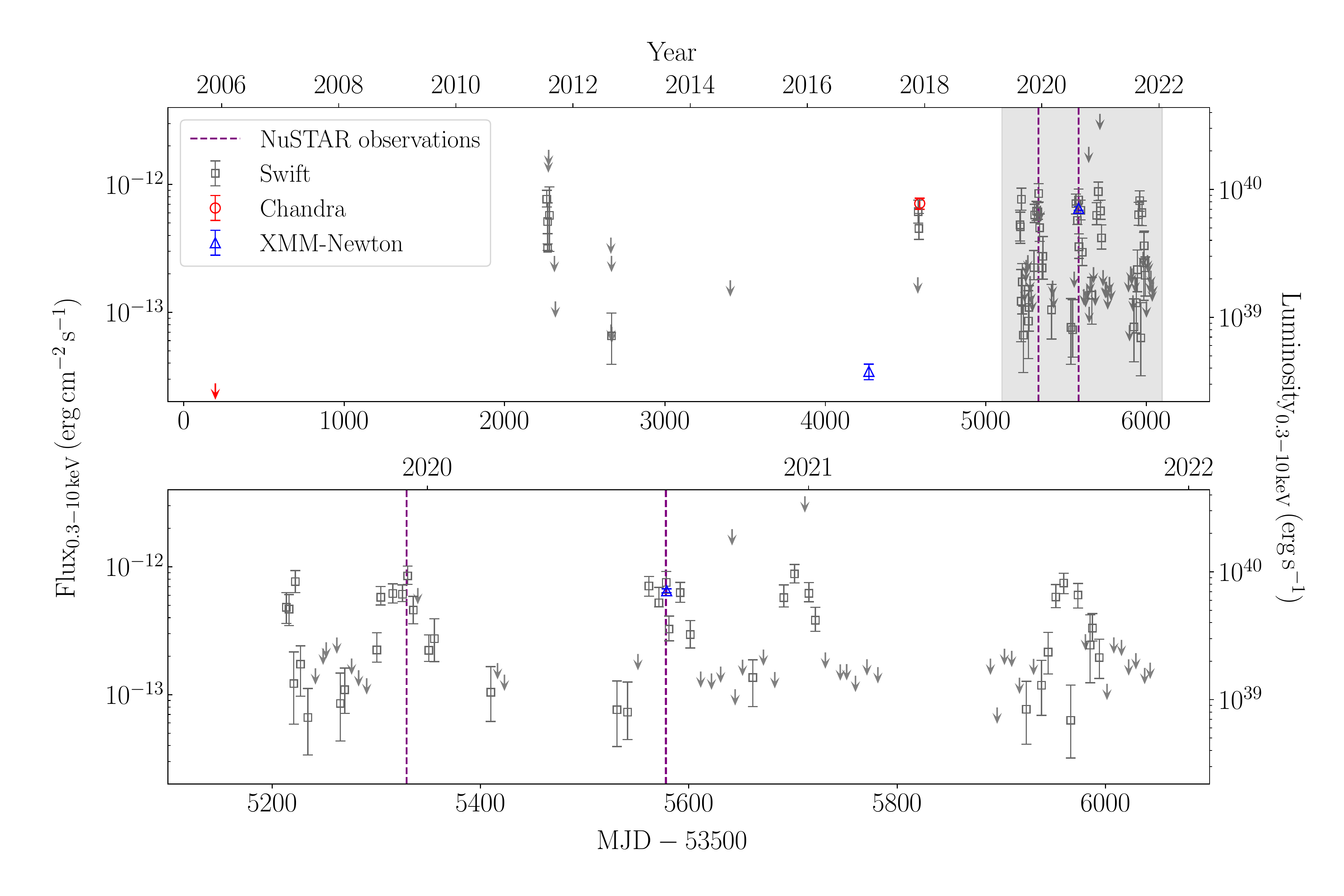}
	\end{center}
	\vspace{-6mm}
	\caption{The long term light curve for NGC~925~ULX-3, showing fluxes and luminosities in the 0.3--10\,keV energy band. \swift\ data points (square) and upper limits are shown in grey, \chandra\ data points (circle) and upper limits are shown in red, and \xmm\ data points (triangle) are shown in blue. The times of \nustar\ observations are indicated by purple dashed lines. The upper plot shows all data, starting with the 2005 \chandra\ observation, with the shaded period MJD 58600--59600 shown on the bottom plot zoomed in on the 2019--2021 monitoring program. \label{fig:longtermlc}}
	\vspace{2mm}
\end{figure*}

\subsection{XMM-Newton} \label{sec:xmm}

We extracted the data from \xmm\ observation 0862760201 from the EPIC-pn and EPIC-MOS instruments using the \xmm\ SAS v18.0.0 software. We produced calibrated event lists using the tasks {\tt emproc} and {\tt epproc}. Periods of high background flaring were removed by filtering out intervals of time during which the $>$10\,keV count rate exceeded 0.35\,cts/s across the EPIC-MOS detectors and the 10--12\,keV count rate exceeded 0.5\,cts/s across the EPIC-pn detector. There were no major periods of background flaring during this observation, so this filtering had minimal effect. We extracted data products from a 30$^{\prime\prime}$ radius circular source region, using a 45$^{\prime\prime}$ radius circular region on the same chip with a similar distance from the readout node for the background. Events with {\tt FLAG==0 \&\& PATTERN$<$4} were selected from EPIC-pn, and {\tt PATTERN$<$12} from EPIC-MOS. The EPIC-pn light curve was extracted with a bin size equal to the pn time resolution, i.e. 73.4\,ms. The spectrum was grouped into 20 counts per bin to allow for Gaussian statistics when fitting, and oversampling set to three times the spectral resolution. The redistribution matrices and auxiliary response files were created using the tasks {\tt rmfgen} and {\tt arfgen}, respectively. 

The 0.3--10\,keV flux for each observation was calculated from the best-fitting absorbed power-law model to the \xmm\ data by itself, using $N_{\rm H} = 1.4\times10^{21}$\,cm$^{-2}$ as the fixed absorption for the low-flux archival observation. 

\subsection{NuSTAR} \label{sec:nustar}

The two \nustar\ observations were reduced using NuSTARDAS v2.0.0, with CALDB version 20211020. The {\tt nupipeline} routine was used to produce clean event files, and {\tt nuproducts} used to extract source and background spectra and response files. A 30$^{\prime\prime}$ radius source region at the \chandra\ source location and centered on the source PSF was used. This is smaller than often typical for \nustar\ extraction regions to minimize contamination from the nearby source ULX\=/2, which is $\sim$50$^{\prime\prime}$ from ULX\=/3. The background spectrum was extracted from a 60$^{\prime\prime}$ region located nearby on the chip but outside of the galaxy. Spectra were grouped into 20 counts per bin, as for \xmm. 

\section{Results} \label{sec:results}

We show the long-term light curve of ULX-3, both over the course of all X-ray observations and zoomed in on our monitoring campaign, in Fig.~\ref{fig:longtermlc}. The light curve clearly shows repeated brightening and dimming, with the \swift\ monitoring campaigns covering four complete excursions into a bright state, plus the latter half of a brightening at the beginning of the 2019 monitoring. Each bright state reaches a reasonably consistent peak flux around $8\times10^{-13}$\,\fluxcgs\ ($L \sim 9\times10^{39}$\,\lumcgs), with the highest flux we detect during the monitoring being $1.2\pm0.3 \times10^{-12}$\,\fluxcgs\ ($L = 1.3\pm0.3 \times10^{40}$\,\lumcgs). Outside of these bright excursions, the source is not always reliably detected by \swift, but upper limits or detections tend to be consistent with a flux of $\sim$1$\times10^{-13}$\,\fluxcgs\ or a little lower. The lowest-flux detection was by \xmm\ in 2017, at a flux of $3.5\pm0.5 \times10^{-14}$\,\fluxcgs\ ($L = 3.8\pm0.5 \times10^{38}$\,\lumcgs). Even lower than this was an upper limit to the flux found by \chandra\ in 2005 at $2.8\times10^{-14}$\,\fluxcgs. The full extent of the observed variability amplitude between the highest-flux detection and the lowest upper limit is a factor of $\gtrsim$40.

\begin{figure*}
	\begin{center}
	\includegraphics[width=16cm]{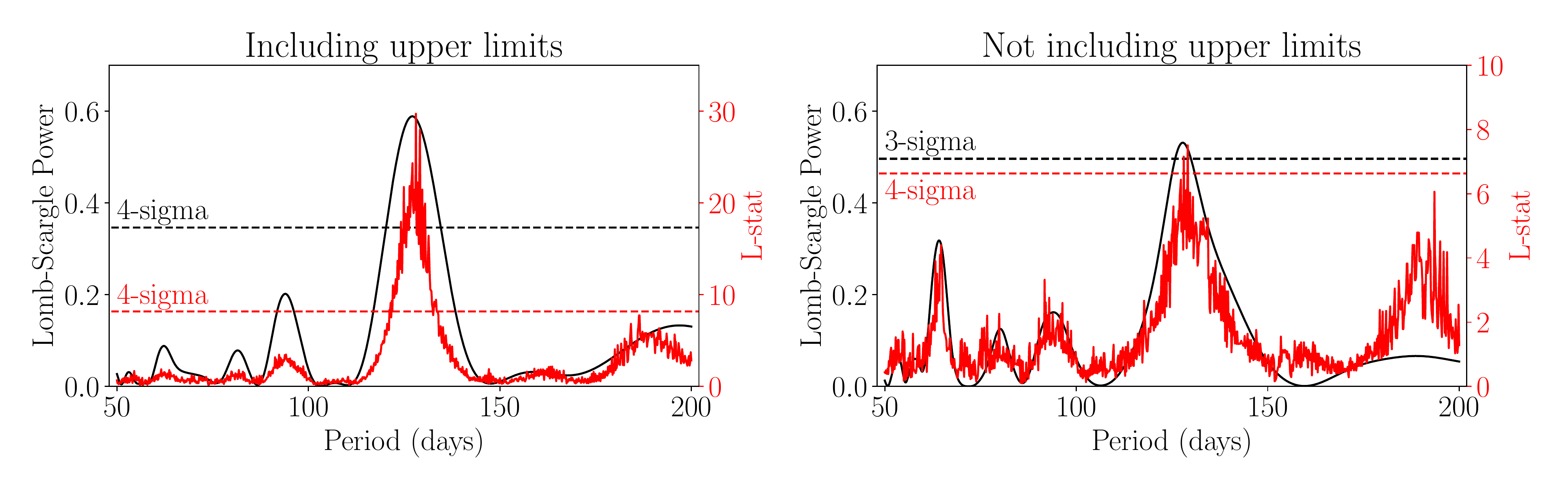}
	\end{center}
	\vspace{-6mm}
	\caption{The value of the Lomb-Scargle power (black) and the epoch folding L-statistic (red) by period for all low-energy data over the regular \swift\ monitoring during 2019--2021. The results when upper limits are included are shown on the left, and when they are excluded on the right. 3$\sigma$ or 4$\sigma$ significance levels are indicated with dashed lines. \label{fig:periodresults}}
\end{figure*}

\subsection{Timing Analysis} \label{sec:timing}

To test for long-term periodicity in the light curve, we performed two tests: the Lomb-Scargle test \citep{lomb76,scargle82}, as implemented by {\tt astropy}, and often used for finding signals in sparse or irregularly-sampled data; and an epoch-folding test \citep{leahy83}, for which we calculate the L-statistic \citep{davies90}. For the epoch-folding test, we folded the flux light curve over each period in eight phase bins, to minimize instances of under-full phase bins.

For each method, we tested for periods between 50 and 200 days, in intervals of 0.1 days, for 1500 trial periods in total. A lower limit of 50 days was chosen because while some ULXs show periodicity on a faster timescale than this, we found it was common for ULX-3 to remain bright for $\sim$50 days at a time, so we set this as a lower bound. An upper bound of 200 days was chosen since there are at least four excursions into a high-luminosity regime in the $\sim$800 days of regular \swift\ monitoring. We included both detections and upper limits from all soft X-ray observatories (i.e. \swift, \xmm, and \chandra) in this analysis, treating upper limits as zero-flux detections. (We obtain the same results at very similar significance if we instead use the measured aperture flux for the observations with upper limits. If we use the method in \citealt{salvaggio22} and assign each upper limit a flux between zero and the value of the upper limit, the significance of our detection is much reduced, but we found that this effect is mainly due to two observations with weakly constraining upper limits higher than the maximum flux measured for this source. Picking a random flux in the given range for these observations is unrealistic and distorts the results. If these two upper limits are discounted, the significance becomes once again very similar to the zero-flux or aperture-flux treatments of the upper limits). 

We performed these analyses for data points only in the 2019--2021 period in which regular monitoring was undertaken, since it is both possible for super-orbital periodicity to change (e.g. \citealt{trowbridge07,brightman22}) or disappear (e.g. \citealt{grise13}) and for superorbital periodicity to remain consistent even through periods when the flux drops dramatically for other reasons (e.g. \citealt{fuerst17}). In each case, the significance of a periodicity was estimated by performing Monte-Carlo simulations following \citet{walton16}, in which we simulated 10,000 light curves based on a red noise power spectrum at 20 times the total duration of observations and 2\,ks resolution (a typical duration for a \swift\ observation). These light curves were then sampled at a similar distribution of times to the observations, with a scatter of $\pm$1 day applied, and the Lomb-Scargle and epoch-folding tests performed on each. We plot the test statistic results by period, as well as the 4/3$\sigma$ level where 99.994\%/99.7\% of the simulated test statistics lie beneath this threshold (Fig.~\ref{fig:periodresults}, left). As a check, we also performed this analysis on the detections by themselves, leaving out upper limits. We find that we recover approximately the same period, albeit at a lower significance due to the lower number of data points (Fig.~\ref{fig:periodresults}, right). 

\begin{figure}
	\begin{center}
	\includegraphics[width=8cm]{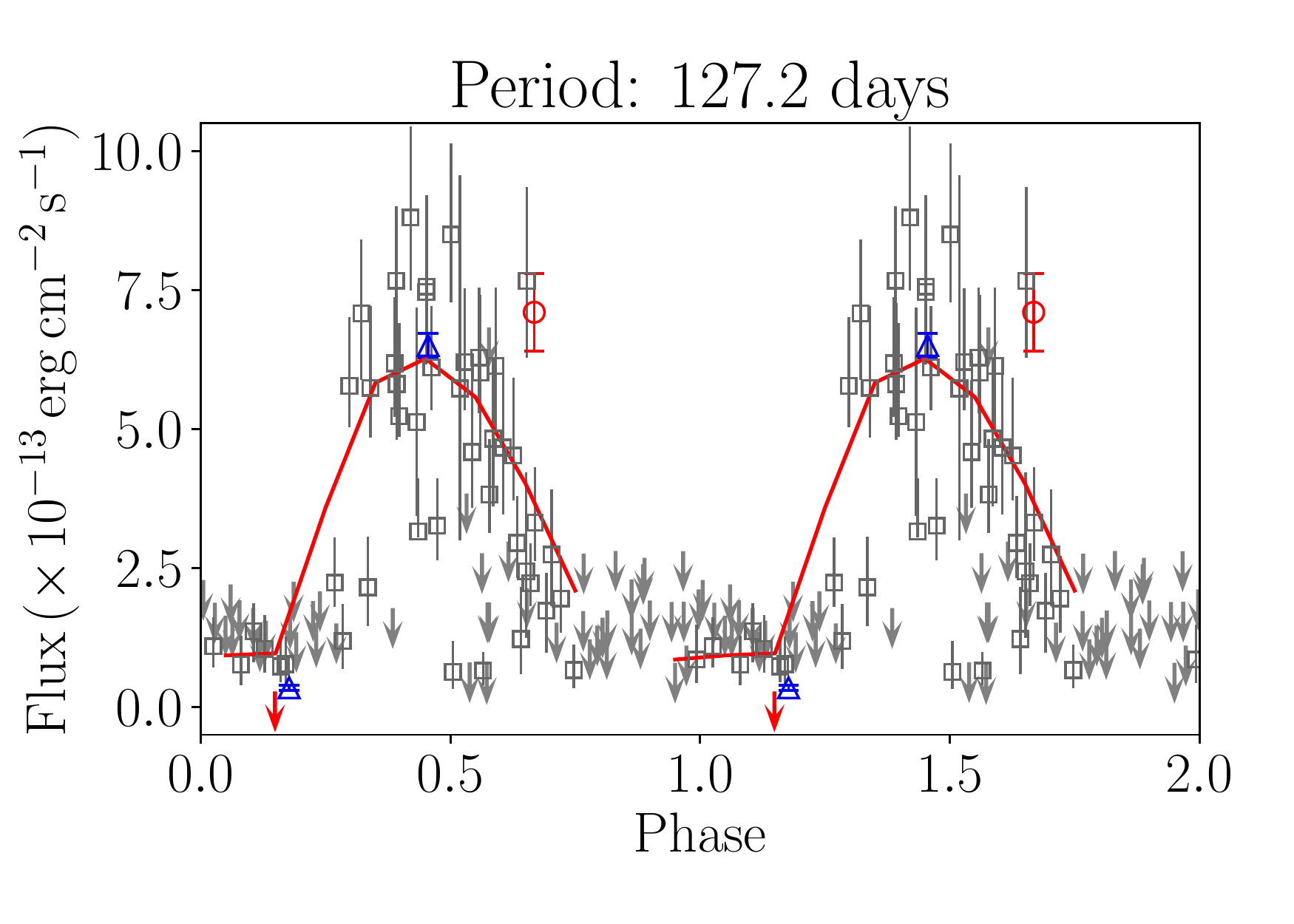}
	\vspace{-4mm}
	\includegraphics[width=8cm]{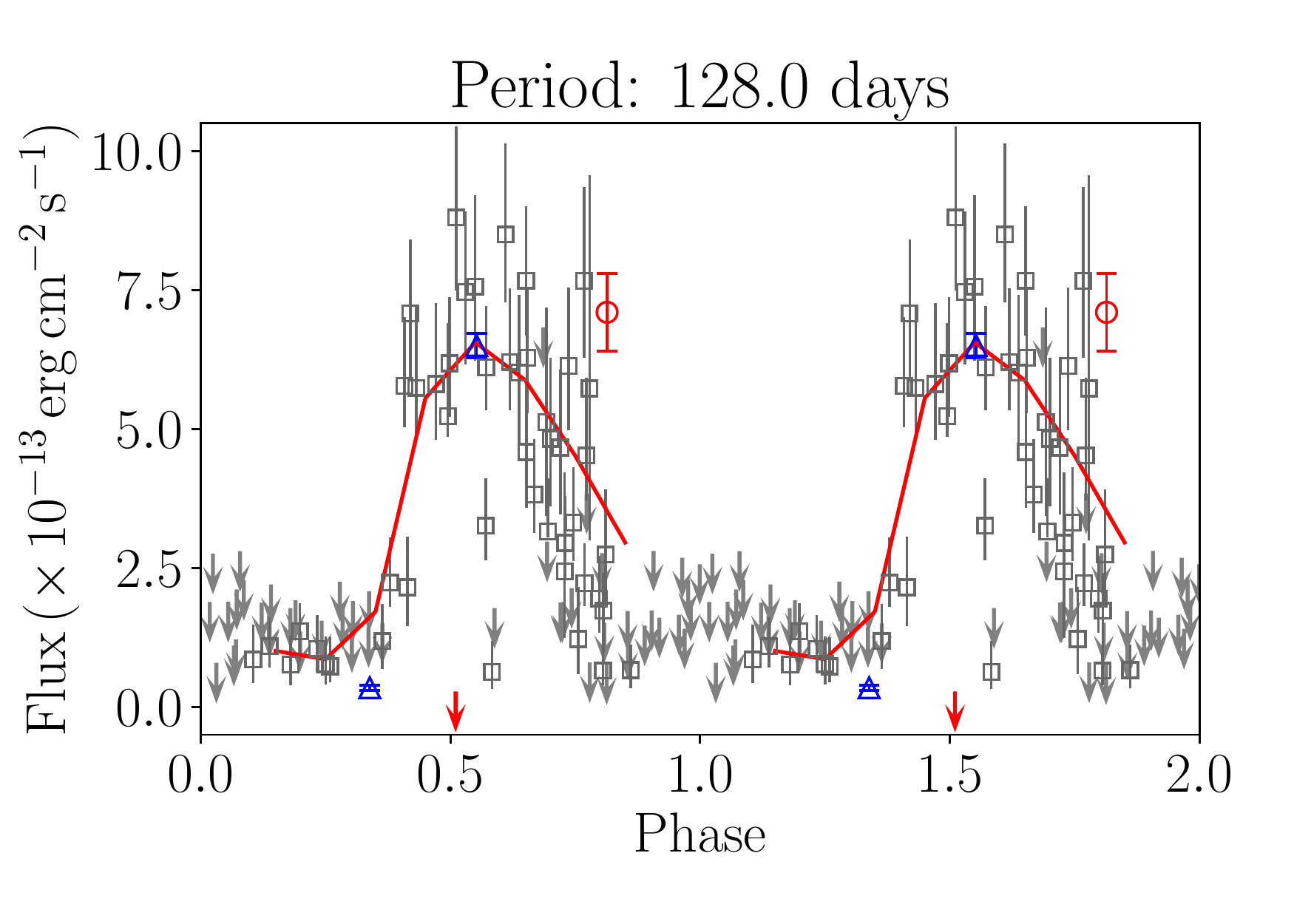}
	\end{center}
	\vspace{-4mm}
	\caption{The X-ray light curve folded over a 127.2-day (top) and 128-day (bottom) periodicity. Two phase cycles are shown for clarity. Symbols are as in Fig.~\ref{fig:longtermlc}. The weighted mean of the detections in each 0.1 phase bin is shown in red -- where there is a gap, these are phase bins only containing upper limits, and so a mean flux could not be found. \label{fig:phasefolded}}
\end{figure}

\begin{figure*}
	\begin{center}
	\includegraphics[width=18cm]{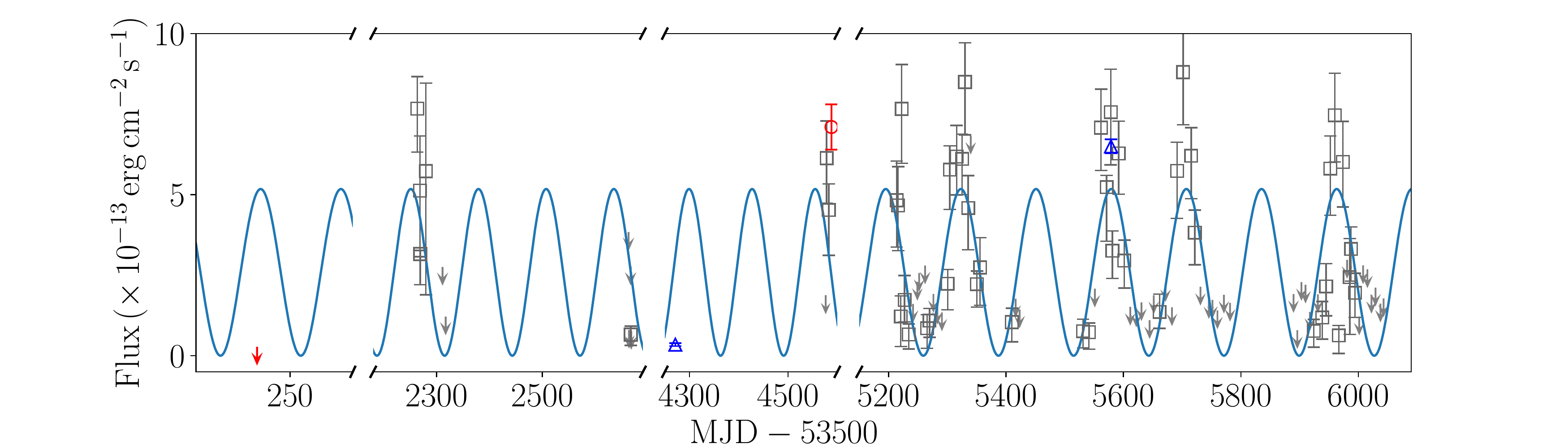}
	\end{center}
	\vspace{-4mm}
	\caption{A 128-day period sinusoid (blue) fitted to the 2019--2021 monitoring data and extrapolated through to archival observations. Symbols are as in Fig.~\ref{fig:longtermlc}. \label{fig:lcsine}}
\end{figure*}


To find the error on the resulting periods, we next simulated 1,000 light curves in a similar fashion as above, but this time using as a model the best-fitting sinusoid to the data, with the period set to that found by each test. They were sampled in the same manner and both tests used to find the period for each light curve. We used the distribution of the results from these simulations to determine the 90\% confidence interval.

We find that the Lomb-Scargle and epoch-folding tests give consistent results, of $127.2\pm0.4$ and $128\pm1$ days respectively. We show phase-folded plots for both of these periods in Fig.~\ref{fig:phasefolded}. These results are consistent with and confirm the detection of a $126\pm2$ day period first reported by \citet{salvaggio22}. If we extrapolate these periods back to past data, we find that most previous data points are roughly consistent with such a period, except for the first \chandra\ non-detection in the case of a slightly longer period (Fig.~\ref{fig:lcsine}).

We also searched for pulsations in both the \xmm\ and \nustar\ data using {\tt HENaccelsearch} from the HENDRICS v5 software package \citep{bachetti18}, searching in the 0.01–5\,Hz frequency range and using an accelerated search since ULX pulsars tend to have such dramatic spin-up that a non-accelerated periodicity search would fail to detect pulsations. We did not find any significant detections or potential candidates. Since no tentative signals were apparent from an accelerated pulsation search, we did not perform any more complex analysis involving orbital period modulations, although we note that if the 127--128-day periodicity is an orbital period, we would not expect it to have a dramatic effect on the pulse period modulation over the course of our observation in addition to spin-up. See Section~\ref{sec:discperiod} for further discussion of the long-term periodicity.

We used the {\tt stingray} software package to simulate observed light curves of a pulsation at various pulse fractions, following the method described in \citet{fuerst21}, and found that we can place an upper limit on the pulsed fraction of $\sim$40\% in the \xmm\ band for pulse periods from 0.01\,Hz to $\sim$2\,Hz, with the upper limit quickly increasing to 100\% for higher pulse frequencies approaching the EPIC-pn timing resolution. There are insufficient data to place a meaningful limit on the pulsed fraction in the \nustar\ band.

\subsection{Spectral Analysis} \label{sec:spectra}

We used XSPEC v12.10 \citep{arnaud96} to perform all spectral fitting, and all quoted models are given in XSPEC syntax. In all cases, spectra are grouped into at least 20 counts per bin to allow for $\chi^2$ statistics to be used in fitting. We give uncertainties at the 90\% confidence level, and we use the abundance tables of \citet{wilms00} throughout. We use only a single {\tt tbabs} model to account for absorption due to the interstellar medium, though we note that the Galactic contribution to this is $N_{\rm H} = 7.26\times10^{20}$\,cm$^{-2}$ \citep{willingale13}. We considered \swift\ and \xmm\ data in the energy range 0.3--10\,keV, and \nustar\ data in the range 3--20\,keV, above which the background dominates the spectrum for this source. We show all spectral fitting results in Table~\ref{tab:fitresults}.

\subsubsection{\nustar\ Epoch 1} \label{sec:swiftnustar}

The first \nustar\ DDT observation was taken when the source was not visible to \xmm, so the low-energy coverage of the spectrum is provided by \swift. In order to increase the signal at low energies, we used the FTOOLS routine {\tt addascaspec} to co-add the observation taken simultaneously with the \nustar\ observation (observation ID: 00089002001) with the three preceding \swift\ observations which were all consistent in flux (observation IDs: 00045596035--37). Since there are insufficient low-energy data to place good constraints on $N_{\rm H}$, we froze it to $N_{\rm H} = 1.4\times10^{21}$\,cm$^{-2}$, based on our fits to the bright \xmm\ detection (see Section~\ref{sec:xmmnustar}).

\begin{figure}
	\begin{center}
	\includegraphics[width=8.5cm]{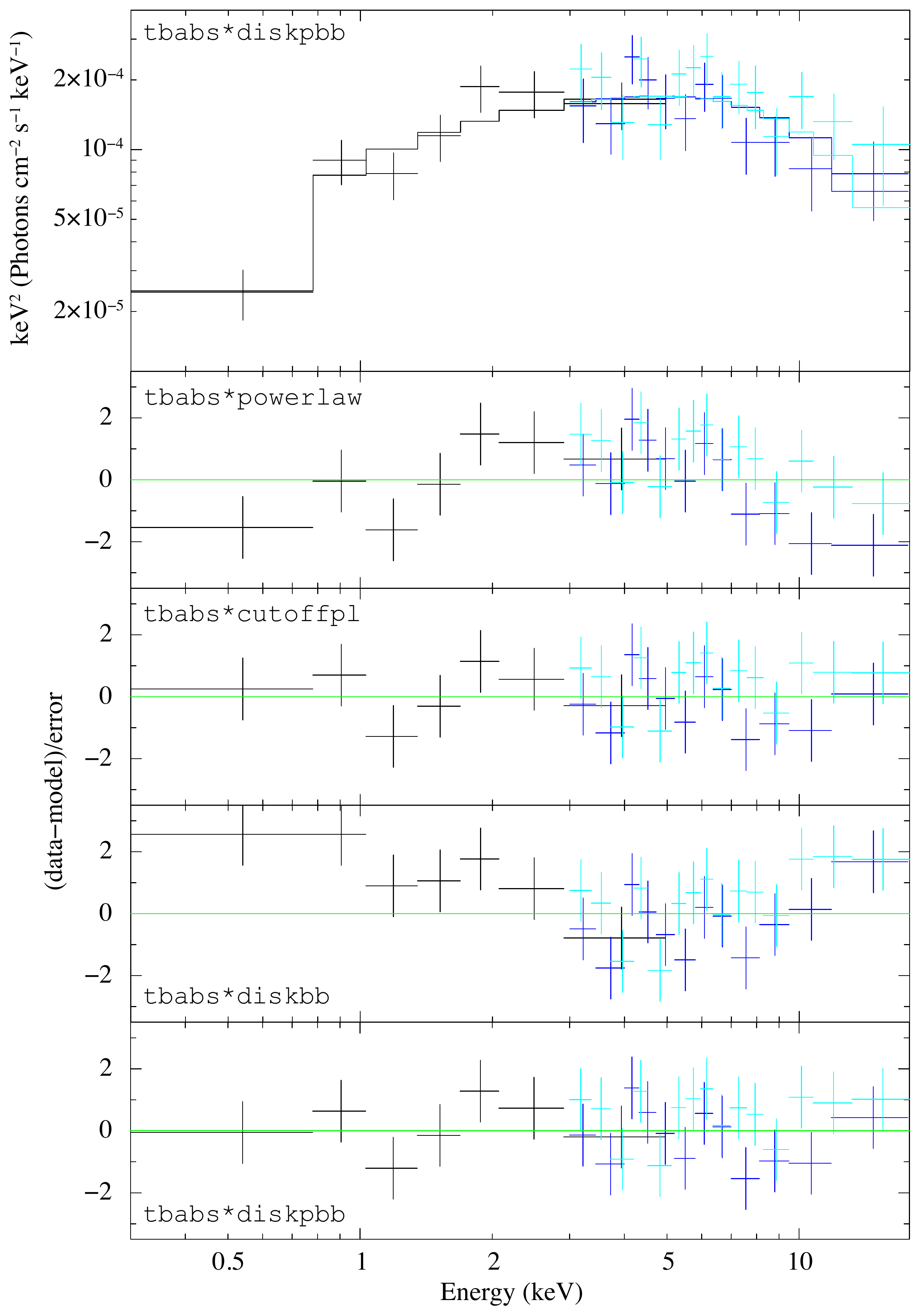}
	\end{center}
	\vspace{-4mm}
	\caption{The unfolded \swift\ (black) and \nustar\ FPMA and FPMB (blue and cyan respectively) spectra fitted with the best-fitting {\tt tbabs*diskpbb} model, and the residuals for all fitted models. \label{fig:toospectraplot}}
\end{figure}

\begin{figure}
	\begin{center}
	\includegraphics[width=8.5cm]{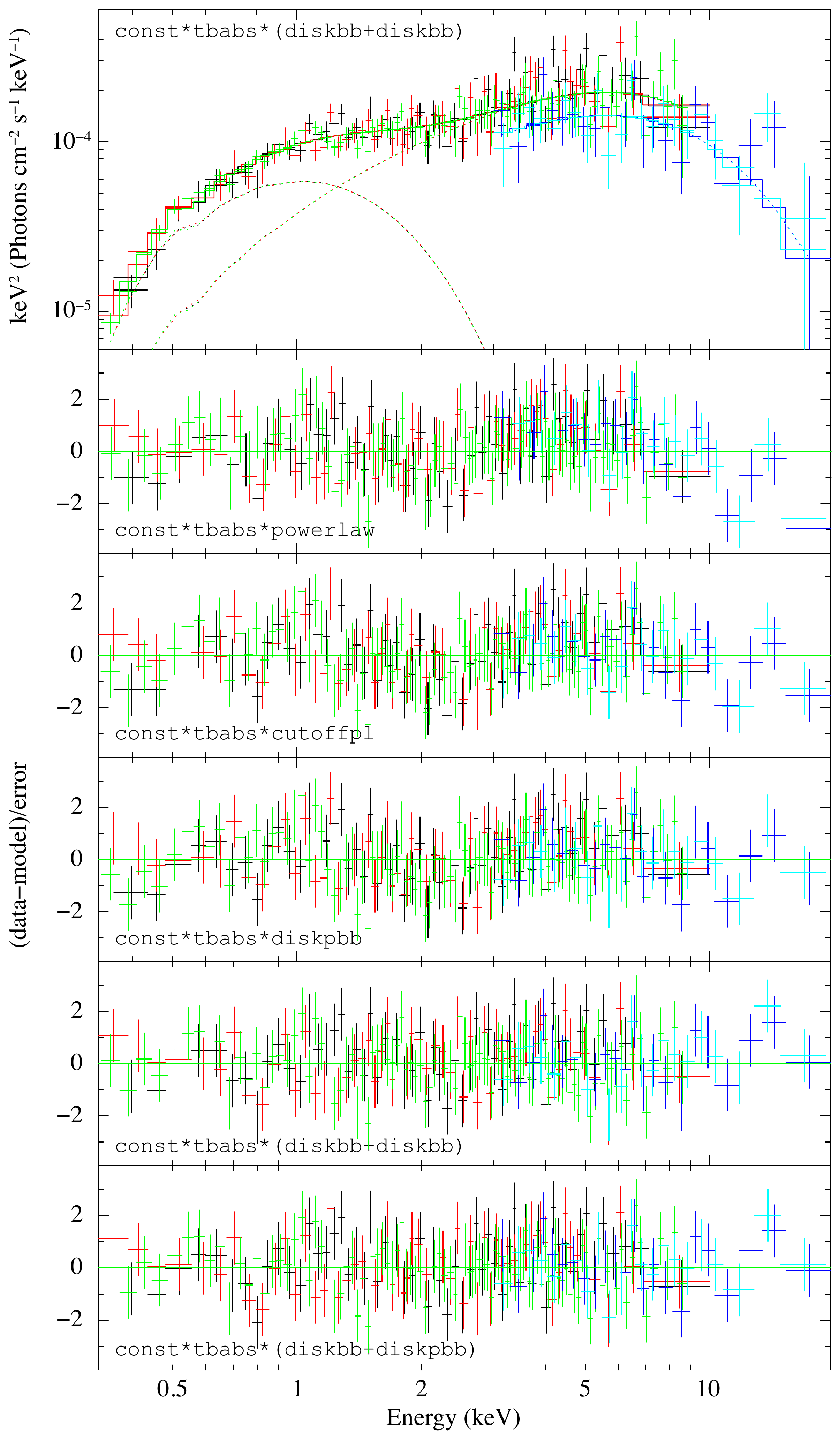}
	\end{center}
	\vspace{-4mm}
	\caption{The unfolded \xmm\ MOS1, MOS2, and pn spectra (black, red, and green respectively) and \nustar\ FPMA and FPMB (blue and cyan respectively) spectra fitted with the {\tt const*tbabs*(diskbb+diskbb)} model, and the residuals for all fitted models. \label{fig:xmmnuspectraplot}}
\end{figure}

\begin{deluxetable*}{@{~}c@{~}c@{~}c@{~}c@{~}c@{~}c@{~}c@{~}c@{~}c@{~}c@{~}c@{~}}
	\tablecaption{The spectral fitting results. \label{tab:fitresults}} 
	\tablecolumns{11}
	\tablenum{2}
	\tablewidth{0pt}
	\tablehead{
		 \colhead{Model\tablenotemark{a}} & \colhead{Constant} & \colhead{$N_{\rm H}$} & \colhead{$\Gamma$} & \colhead{E$_{\rm cut}$} & \colhead{T$_{\rm in,1}$} & \colhead{T$_{\rm in,2}$} & \colhead{$p$} & \colhead{norm$_1$} & \colhead{norm$_2$} & \colhead{$\chi^2/{\rm dof}$} \\
		 \colhead{\tt tbabs*} & \colhead{} & \colhead{$\times10^{21}\,{\rm cm}^2$} & \colhead{} & \colhead{keV} & \colhead{keV} & \colhead{keV} & \colhead{} & \colhead{$\times10^{-4}$} & \colhead{$\times10^{-4}$} & \colhead{}
	}
	\startdata
	\hline
	\multicolumn{11}{c}{Co-added \swift\ observations + \nustar\ 90501351002} \\
	\hline
	{\tt pl} & - & $1.4$\tablenotemark{b} & $1.96\pm0.08$ & - & - & - & - & $1.3\pm0.2$ & - & 46.3/32 \\
	{\tt cpl} & - & $1.4$ & $1.3\pm0.3$ & $6^{+4}_{-2}$ & - & - & - & $1.3\pm0.2$ & - & 27.3/31 \\
	{\tt dbb} & - & $1.4$ & - & - & $2.0\pm0.2$ & - & - & $20^{+12}_{-8}$ & - & 50.9/32 \\
	{\tt dpbb} & - & $1.4$ & - & - & $3.0^{+0.9}_{-0.5}$ & - & $0.55^{+0.04}_{-0.03}$ & $<2.6$ & - & 26.2/31 \\
	{\tt dbb+dbb} & - & $1.4$ & - & - & $0.8\pm0.4$ & $2.8^{+2.2}_{-0.7}$ & - & $400^{+6800}_{-300}$ & $4^{+9}_{-3}$ & 28.0/30 \\
	\hline
	\multicolumn{11}{c}{\xmm\ 0862760201 only} \\
	\hline
	{\tt pl} & - & $1.4\pm0.2$ & $1.74\pm0.05$ & - & - & - & - & $1.15\pm0.05$ & - & 236.3/217 \\
	\hline
	\multicolumn{11}{c}{\xmm\ 0862760201 + \nustar\ 80601305002} \\
	\hline
	{\tt c*pl} & $0.63\pm0.07$ & $1.4\pm0.1$ & $1.78\pm0.05$ & - & - & - & - & $1.19\pm0.05$ & - & 306.3/261 \\
	{\tt c*cpl} & $0.74\pm0.08$ & $1.1\pm0.2$ & $1.5\pm0.1$ & $12^{+8}_{-4}$ & - & - & - & $1.19\pm0.05$ & - & 288.0/260 \\
	{\tt c*dpbb} & $0.76\pm0.08$ & $1.2\pm0.2$ & - & - & $3.7^{+0.7}_{-0.5}$ & - & $0.55\pm0.01$ & $0.6^{+0.4}_{-0.3}$ & - & 279.5/260 \\
	{\tt c*(dbb+dbb)} & $0.73^{+0.08}_{-0.07}$ & $1.1^{+0.3}_{-0.2}$ & - & - & $0.35\pm0.04$ & $2.4\pm0.2$ & - & $700^{+500}_{-300}$ & $9^{+3}_{-2}$ & 248.7/259 \\
	{\tt c*(dbb+dpbb)} & $0.73^{+0.08}_{-0.07}$ & $1.2^{+0.4}_{-0.3}$ & - & - & $0.32\pm0.07$ & $2.6^{+0.6}_{-0.5}$ & $0.67^{+0.21}_{-0.07}$ & $900^{+1400}_{-400}$ & $5^{+13}_{-4}$ & 247.9/258 \\
	\hline
	\multicolumn{11}{c}{Low-luminosity \swift\ observations} \\
	\hline
	{\tt pl} & - & $1.4$ & $1.8\pm0.3$ & - & - & - & - & $0.12\pm0.02$ & - & 6.4/9 \\
	\hline
	\enddata
	\tablenotetext{a}{The model names are abbreviated as follows: {\tt pl} = {\tt powerlaw}, {\tt cpl} = {\tt cutoffpl}, {\tt dbb} = {\tt diskbb}, {\tt dpbb} = {\tt diskpbb}.}
	\tablenotetext{b}{Value frozen due to the low-energy data being of insufficient quality to constrain $N_{\rm H}$.}
\end{deluxetable*}

We fitted the spectrum with a number of absorbed single-component models (see Fig.~\ref{fig:toospectraplot} for the best-fitting spectrum and residuals). While a power-law model formally provides an acceptable fit, we find that a cut-off power-law model with a characteristic cut-off energy of $\sim$6\,keV offers a statistically significant improvement ($\Delta\chi^2 = 19$ for 1 degree of freedom). We can statistically rule out a standard disk blackbody model, though a broadened disk model provides an equally acceptable fit to the cut-off power-law. Using the best-fitting model, we find an absorbed source flux of $5.0\pm0.4 \times10^{-13}$\,\fluxcgs\ ($L = 5.5\pm0.4 \times10^{39}$\,\lumcgs) in the 0.3--10\,keV band. For the purposes of comparison, we also fit the spectrum with two disk blackbody models, although we are unable to place strong constraints on the model parameters.

\subsubsection{\nustar\ Epoch 2} \label{sec:xmmnustar}

We began by fitting the \xmm\ spectrum by itself with an absorbed power-law in order to characterize the low-energy emission and quantify the interstellar absorption for use when analyzing the \swift\ data. We found an absorption of $N_{\rm H} = 1.4\times10^{21}$\,cm$^2$. While it is possible for $N_{\rm H}$ to change between observations, it is nonetheless likely to be a better approximation to the true $N_{\rm H}$ than the Galactic value by itself. We found that a simple absorbed power-law model offers a statistically acceptable fit to the low-energy data, although there is an `m'-shaped curvature to the residuals (often seen in ULXs when the low-energy data is fitted with a power-law) that suggests the contribution of multiple components. 

We next fitted the \xmm\ and \nustar\ data simultaneously (see Fig.~\ref{fig:xmmnuspectraplot} for the spectrum and residuals). We included a multiplicative constant in this model fit to account for calibration differences between \xmm\ and \nustar, freezing the value to 1 for \xmm\ and letting it vary for \nustar. This constant is anomalously low, with \xmm\ and \nustar\ expected to be consistent to within $\sim$10\%. The relative normalization may have been affected by the source being very close to one of the \nustar\ chip gaps. 

We found that a cut-off power-law offers a significant improvement in fit over a simple power-law, with a slightly higher cut-off energy although still within the uncertainties of that found for the first \nustar\ epoch. A broadened disk model also offers a statistically acceptable fit. However, in all of these single-component cases, the residuals show a characteristic `m'-shaped curvature indicating that there are likely two components to the emission instead of just one.

We used two-thermal-component models to fit the data, first using two multicolor disk blackbody models, then replacing the higher energy model with a $p$-free broadened disk model, as is often required for ULX spectra (e.g. \citealt{walton18}). We found that both models fit the data well, although using a broadened disk for the hot component over a standard multicolor disk does not show strong evidence for significant broadening and does not provide a statistically significant improvement to the fit. Using the {\tt diskbb+diskbb} model, we calculated an absorbed source flux of $5.5\pm0.1 \times10^{-13}$\,\fluxcgs\ ($L = 6.0\pm0.1 \times10^{39}$\,\lumcgs) in the 0.3--10\,keV band.

\subsubsection{Low Luminosity \swift\ data} \label{sec:lowlumswift}

In order to investigate the spectrum of the source at lower fluxes, we used {\tt addascaspec} to co-add all \swift\ observations with detections below a flux of $2\times10^{-13}$\,\fluxcgs, or with upper limits, discounting low-flux observations or non-detections that are evidently still within an overall bright state (such as the single anomalously low-flux \swift\ observation in the middle of the most recent bright state). In total, we added 60 observations, covering approximately half of the phase cycle (see Fig.~\ref{fig:phasefolded}). As with the previous \swift\ analysis, we froze $N_{\rm H}$ to the value measured for the \xmm\ observation, then we fitted the spectrum with a power-law model. We found that the spectrum could be fitted with a photon index of $\Gamma=1.8\pm0.3$, with an absorbed flux of $6\pm2 \times10^{-14}$\,\fluxcgs\ ($L = 7\pm2 \times10^{38}$\,\lumcgs).

\section{Discussion} \label{sec:disc}

\subsection{A long-term periodicity} \label{sec:discperiod}

We confirmed the high levels of variability of NGC~925 ULX-3 first reported in E20, and discovered a 127--128-day periodicity over which the source enters and leaves the ULX luminosity regime. This period is consistent within error with that also reported by \citet{salvaggio22}. A small number of ULXs, including several of those that have been identified as neutron star accretors, exhibit (likely) superorbital periodicity on the order of tens to hundreds of days (e.g. \citealt{strohmayer09,lin15,walton16,brightman19}), although we note that NGC~7793~P13 has been suggested to have a $\sim$1500-day superorbital period \citep{motch14,fuerst18}. We also note that Galactic source SS~433, likely also a super-Eddington accreting source viewed at high inclination, exhibits a 164-day periodicity \citep{abell79}. In this context, the NGC~925~ULX-3 periodicity is fairly typical of the ULX population exhibiting such periods. The flux variation of around an order of magnitude is also within the range of such ULX periodicity discovered so far.


In a couple of instances, the source exhibits a dip in luminosity in the middle of an otherwise bright state, the most obvious of these being about halfway through the most recent bright period, during which there is a \swift\ detection consistent in flux with measurements made during low-flux intervals. The phase-folded light curves indicate that this dip may recur at a similar phase of the cycle, and likely does not last longer than a few days, although higher-cadence observations would be required to confirm this. This dipping behavior has also been seen in other ULXs with long periods (e.g. \citealt{pasham13,walton16}), and may be due to periodic/superperiodic obscuration of the source. However, the spectrum of NGC~925~ULX-3 indicates that it is unlikely that we are viewing the source at a high inclination (see Section~\ref{sec:discspectra}). 

This period is more-or-less consistent with archival data points, with the main outlier being the \chandra\ non-detection in 2005. This may be due to a change or disappearance of the super-orbital period in the intervening time, as has been observed elsewhere (e.g. \citealt{grise13,lin15,weng18,brightman22}). However, we note that the \chandra\ upper limit is $2.8\times10^{-14}$\,\fluxcgs, significantly lower than the low-flux detections or our composite low-flux spectrum. This may instead indicate an interval of lower flux by some other mechanism, over which a superorbital period may still persist (e.g. \citealt{fuerst21}). Therefore, while \citet{salvaggio22} claim that the period or phase may have changed in archival observations compared with the periodicity detected in our monitoring campaign, we believe that there is insufficient archival evidence to make such a claim. 

There have been various models proposed to explain superorbital periods in ULXs. While an in-depth theoretical exploration is beyond the scope of this paper, we investigated the implications of a 127--128-day period using some of these models and the assumptions within the corresponding papers. 

\citet{mushtukov17} propose that such variability may be caused by superorbital precession of the magnetic dipole of the accreting neutron star, in the context of a system in which the hot thermal emission originates from an accretion curtain that envelops the entire magnetosphere of the ULX, and the cooler thermal emission comes from a supercritical accretion disk. We find that, if we assume a 1--10\,s pulsation period typical of ULX pulsars so far discovered, a very high magnetic field strength of $>$10$^{15}$\,G is required to produce a superorbital period in the region of 127--128 days, and the corresponding expected temperatures of the thermal components are far lower than we observe ($\sim$0.01\,keV). If we assume a pulsation period on the order of minutes rather than seconds, the model requires lower values of the magnetic field strength of $10^{11}$--$10^{12}$\,G, which gives more reasonable thermal component temperatures in the region of $\sim$1\,keV, although the two model components are always much closer in temperature to each other than what we observe. 

Another proposed model for superorbital variability in ULXs is that of Lense-Thirring precession of the outflowing wind \citep{middleton18,middleton19}, which can be used to construct a timing-accretion plane that may indicate whether the accreting object is a candidate black hole. Here we use the model of a supercritical inner disk and a cooler outer disk, with the observed temperature of the cooler component corresponding to the temperature at the spherization radius $T_{\rm sph}$. In this case, a precession period of 127 days and $T_{\rm sph}=0.35$\,keV place the source comfortably within the region of the timing-accretion plane that contains both black hole and neutron star accretors (see Fig. 4 of \citealt{middleton19}). We find that, using this model, a precession period of 127--128 days can be generated with a magnetic field strength in the range $10^{11}$--$10^{12}$\,G, with the energy fraction used to launch the wind $\epsilon_{\rm wind} = 0.3$--0.35. 

\citet{vasilopoulos20} suggest that superorbital periodicity may instead be due to free precession of the neutron star itself as described in \citet{jones01}, in which the distortion of the neutron star can be derived from its spin and precession period and related to the surface magnetic field, accounting for superconductivity in the neutron star interior \citep{lander14}, with the precession of the accretion disk synchronized to the neutron star via some coupling with the magnetospheric field lines. If we once again assume a spin period of 1--10\,s, a magnetic field of $10^{12}$--$10^{13}$\,G is required for this model to reproduce long-term periodicity on the timescale we measure. Weaker magnetic fields would correspond to shorter spin periods.

A detection of pulsations from this source, allowing the determination of the neutron star spin and any spin-up, would help to further test and constrain these various models. We note, however, that these models assume that the magnetic field is dominated by a dipole component -- it has been suggested that a significant multipolar component to the magnetic field may be present in at least some ULXs \citep{israel17a,tsygankov17}, which may introduce further complexity to these scenarios. We also note that the precession of a radiation-driven warped disk, as seen in some other X-ray binaries \citep{ogilvie01}, has also been proposed as a mechanism for producing a superorbital period for some ULXs (e.g. \citealt{kong16}).

Since an orbital period has not been identified for this source, it is possible that this periodicity is instead orbital rather than superorbital. Several of the ULXs with confirmed orbital periods have periods on the order of days (e.g. \citealt{bachetti14,israel17a}), though NGC 7793 P13 has an orbital period of $\sim$65 days \citep{fuerst21}, so a period on this timescale is not out of the question. Orbital periods on the order of $\sim$100 days have been observed in some Be X-ray binaries (BeXRBs; \citealt{walter15}), with luminosities in the ULX regime for some of the brightest outbursts of BeXRBs, and very large variation in flux being common. However, even during the lowest-flux parts of its phase cycle, NGC~925~ULX-3 has luminosity $>$10$^{38}$\,\lumcgs, far more luminous even than most BeXRB outbursts and certainly more luminous than the intervals between outbursts, for which luminosities of $<$10$^{36}$\,\lumcgs\ are expected. Additionally, BeXRBs tend to have shorter duty cycles due to their eccentric orbits than what we observe for NGC~925~ULX-3, which is close to 50\%. Therefore, it seems more likely that the periodicity we observe is superorbital in nature. 

\subsection{Spectral behavior} \label{sec:discspectra}

The spectrum at higher fluxes is typical of a ULX in the super-Eddington ultraluminous state \citep{gladstone09}, with a turnover in the \nustar\ band and a complex shape that can be fitted using two disk blackbody components. The fact that we see a high-energy emission component indicates that we are unlikely to be viewing the source at a high inclination -- super-Eddington systems at high inclinations tend to show very soft spectra dominated by the cool thermal component due to the hotter central region being obscured (e.g. \citealt{urquhart16}). The broadband spectra of ULXs in the ultraluminous state will often show broadening in the hotter component, associated with an advection-dominated supercritical accretion disk, as well as a steep power-law excess at high energies (e.g. \citealt{bachetti13,mukherjee15,rana15}) which may be due to emission from the accretion column of a neutron star (e.g. \citealt{walton18}). We do not find any strong evidence of broadening or the power-law tail for NGC~925 ULX-3, although this is likely due to the limited data quality of our \nustar\ observations, for which we do not have many data points above 10\,keV. 

The spectra between the two epochs are quite consistent in shape and flux, so it is reasonable to assume that the source is in the same state during the first \nustar\ epoch as it is in the second -- fitting the first epoch with a two-thermal-component model suggests that the cooler of the two thermal components may be hotter than that of the second epoch, though the quality of the \swift\ data is insufficient to draw any strong conclusions from this. Also, since the first \nustar\ epoch is well-fitted with a {\tt diskpbb} model alone, with similar parameters to the hot component of the second epoch, it may be the case that a cool component simply increased in normalization between the two observations.

A temperature for the cooler component of $kT \approx 0.35$\,keV is fairly typical for the ULX population with good broadband data ($0.8$\,keV is unusually hot for this component, but our measurement for the first \nustar\ epoch is consistent within measurement uncertainties with more typical cooler values). For the hotter component, $kT = 2.4$--2.8\,keV is also reasonable, if on the high end, compared with other known ULXs. If this component originates from a supercritical accretion disk between the spherization radius $r_{\rm sph}$ and an inner radius truncated at the magnetospheric radius $r_{\rm m}$ of an accreting neutron star, in the context of the model proposed in \citet{walton18}, hotter temperatures of this component would correspond to a smaller $r_{\rm m}$ (or even an accretion disk that is not truncated at all, should the accretor instead be a black hole). If this were the case, we would expect to see a broadened hot component (in highly-truncated cases with $r_{\rm m}$ close to $r_{\rm sph}$, even a supercritical disk may appear as a narrow spectral component as the emission would originate from a limited band of radii). Higher-quality \nustar\ data would be required to explore the shape of the hot component in greater detail and confirm the presence or absence of broadening in the hot component.

In this model, the cooler component would either originate from an outer thin disk whose inner edge is at $r_{\rm sph}$, or from the radiatively-driven outflow within $r_{\rm sph}$ (given an outer disk's requirement to be sub-Eddington, it is more likely for this component to be the latter for this source, since the luminosity of the cool thermal component alone is $\sim$2.5$\times10^{39}$\,\lumcgs). In either case, the ratio of temperatures between the two thermal components can give some rough indication of the relative sizes of $r_{\rm m}$ and $r_{\rm sph}$. The temperature ratio between the two thermal components of NGC~925~ULX-3 in the second epoch is 7--8, whereas ULX pulsars tend to have lower ratios of $\sim$3 \citep{walton18}. A larger temperature ratio would indicate that $r_{\rm m} << r_{\rm sph}$, with the inner thick disk not severely truncated by the magnetic field and contributing more emission than the pulsed emission coming from the neutron star accretion column. This may contribute to the dilution of pulsations, making them harder to detect. 

Another factor that may contribute to the non-detection of pulsations is geometric beaming by a collimating wind with a large scale height compared to $r_{\rm m}$, in which the amount of photon scattering within the accretion funnel dilutes the intrinsic pulsed signal. This is a scenario that can be ruled out for observed ULX pulsars with high pulsed fractions (e.g. \citealt{mushtukov21}) but conversely may apply to ULXs in which pulsations aren't detected, particularly those which show evidence of a strong outflowing wind component such as in NGC~925~ULX-3. These potential factors indicate that our inability to detect pulsations from this source does not necessarily rule out a neutron star accretor. (We also note that pulsed fractions in the \xmm\ band of 10--20\% have been observed in other ULXs, consistent with our upper limit of 40\%, e.g. \citealt{fuerst16,rodriguezcastillo19}). 

The low-flux spectrum is consistent in shape with the low-flux \xmm\ observation analyzed in E20 (which was well-fitted with a power-law model with $\Gamma = 1.8^{+0.2}_{-0.1}$) as well as with the high-flux \xmm\ spectrum analyzed in this work. Therefore we find no particular evidence of spectral change across the profile of the X-ray period, such as that observed in NGC~5907~ULX\=/1 \citep{fuerst17}, although deeper observations of the low-flux regime for this source will be required to place stronger constraints on the shape of the low-flux spectrum and search for evidence of spectral change with flux. Since there is little spectral change despite the source moving into and out of the super-Eddington luminosity regime, this indicates that there is unlikely to be an intrinsic change in the accretion state itself to produce this variation. The luminosity of $10^{39}$\,\lumcgs\ used to define ULXs comes from the assumption of a $\sim$10\,\msol\ black hole -- all luminosities that we measure for this source are above the Eddington luminosity for a 1.4\,\msol\ neutron star, so there is no requirement for this source to be changing between super-Eddington and sub-Eddington accretion states. \\

\section{Conclusion} \label{sec:conc}

NGC~925~ULX-3 shows the two-component spectrum typical of a super-Eddington accreting system, and demonstrates long-term, likely superorbital variation that is common in ULXs containing neutron star accretors, with a period of 127--128 days. Various different models of superorbital periodicity in ULXs could potentially explain its behavior, with neutron star magnetic fields up to $\sim10^{12}$\,G providing reasonable solutions. While the source luminosity varies by approximately an order of magnitude in flux, there is no indication of a change in accretion state. Therefore we believe this source to be a strong candidate for being a neutron star ULX and a valuable target for follow-up investigations searching for pulsations. 

\acknowledgments

We thank our anonymous referee for useful suggestions to improve this paper. We wish to thank Dan Stern for useful comments on this work. This work was supported by NASA grants 80NSSC20K1496 and 80NSSC21K0873, as well as by NASA contract NNG08FD60C. TPR acknowledges funding from STFC consolidated grant ST/000244/1.  

We wish to thank the \swift\ PI, Brad Cenko, for approving the target of opportunity requests we made to monitor NGC~925. We also wish to thank the \nustar\ PI, Fiona Harrison, for approving the DDT request to observe NGC~925~ULX-3 in December 2019. The scientific results reported in this article are based on observations made by \xmm, an ESA science mission with instruments and contributions directly funded by ESA Member States and NASA; observations by the \nustar\ mission, a project led by the California Institute of Technology, managed by JPL, and funded by NASA; and observations by the \swift\ mission, as well as public data from the \swift\ data archive provided by the UK Swift Science Data Centre at the University of Leicester. We thank the observing teams for these missions for carrying out these observations. 

\vspace{5mm}
\facilities{XMM, CXO, Swift(XRT), NuSTAR}

\software{astropy \citep{astropy13}, CIAO \citep{fruscione06}, FTOOLS \citep{heasarc14}, {\it XMM-Newton} SAS, PIMMS, HENDRICS \citep{bachetti18,stingray}}

\bibliography{ngc925periodpaper}
\bibliographystyle{aasjournal}

\end{document}